\documentclass{cpbtex}
 
\usepackage{graphicx}
\usepackage{bm}
\usepackage{braket}
\usepackage{amsmath}
\usepackage{subcaption}
\usepackage{float}
\usepackage{verbatim}
\usepackage{stfloats}
\usepackage{threeparttable}

\begin{document}


\title{Absolute frequency measurement of molecular iodine hyperfine transitions at 554 nm and its application to stabilize a 369 nm laser for $\mathrm{Yb^+}$ ions cooling\thanks{Project supported by National Natural Science Foundation of China (12073015) and National Key Research and Development Program of China (2021YFA1402100).}}


\author{Y. T. Chen$^{1}$, \ N. C. Xin$^{1}$\thanks{Present address: AVIC Xi'an Flight Automatic Control Research Institute}, \ H. R. Qin$^{2}$\thanks{Present address: 2012 Lab, Huawei Technologies Ltd. Co.},  \ S. N. Miao$^{1}$, \ Y. Zheng$^{2}$, \ J. W. Zhang$^{1}$\thanks{Jian-Wei Zhang. E-mail:~zhangjw@tsinghua.edu.cn}, \ L. J. Wang$^{1,2}$\\
$^{1}$ State Key Laboratory of Precision Measurement Technology and Instruments, \\ Key Laboratory of Photon Measurement and Control Technology of Ministry of Education, \\ Department of Precision Instrument, Tsinghua University \\  
$^{2}$ Department of Physics, Tsinghua University } 


\date{\today}
\maketitle

\begin{abstract}
We investigate 13 hyperfine structures of transition lines of $\mathrm{^{127}I_2}$ near 554 nm, namely, the R(50) 22-0, P(46) 22-0, P(121) 24-0, P(69) 25-1, R(146) 25-0, R(147) 28-1, P(160) 26-0, P(102) 26-1, R(96) 23-0, R(49) 22-0, P(45) 22-0, P(92) 23-0, and R(72) 25-1 transitions, and measure their absolute frequencies with an optical frequency comb. A 369 nm frequency-tripled laser is frequency stabilized by locking the 554 nm harmonic-frequency laser to the R(146) 25-0 $a_{15}$ line of $\mathrm{^{127}I_2}$ via modulation transfer spectroscopy. A frequency stability of $5 \times 10^{-12}$ is observed over a 1000 s integration time. The measurement of the molecular iodine spectroscopy at 554 nm enriches high-precision experimental data, and also enables theoretical predictions. Meanwhile, the 369 nm frequency-tripled laser stabilized by molecular iodine spectroscopy has wide applications in frequency metrology, and quantum information processing based on $\mathrm{Yb^+}$ ions.
\end{abstract}



\section{Introduction}
Frequency stabilized lasers have a wide range of applications in laser cooling and trapping~\cite{1}, quantum frequency standards~\cite{2}, and precision spectroscopy~\cite{3}, etc. The most common method for laser frequency stabilization is via locking the laser frequency to an atomic or molecular absorption line. The rich and broad spectrum of molecular iodine has made it an attractive choice for optical wavelength references and laser frequency stabilization in the visible and near-infrared (NIR) regions~\cite{4}. A laser stabilized via molecular iodine spectroscopy can achieve frequency stabilities in the range of $10^{-14} - 10^{-15}$ over the $1 - 10000$ s time scale~\cite{5}, and the frequency stabilization system is simple and cost-effective. There have been many previous reports on the hyperfine transitions of molecular iodine at different wavelengths: 502 nm~\cite{6}, 515 nm~\cite{7,8}, 531 nm~\cite{9}, 532 nm~\cite{10,11}, 534 nm~\cite{12}, 535 nm~\cite{13}, 543 nm~\cite{14}, 548 nm~\cite{15}, 560 nm~\cite{16}, 561 nm~\cite{17}, 565 nm~\cite{18}, 576 nm~\cite{18}, 578 nm~\cite{19,20}, 585 nm~\cite{18}, 633 nm~\cite{21,22}, 637 nm~\cite{23}, 647 nm~\cite{24}, 671 nm~\cite{25}, 735 nm, and 772 nm~\cite{26}. Based on these experimental results, widely used interpolation formulas for $^{127}\mathrm{I_2}$ hyperfine transition lines have been derived. They can cover the wavelength range from 500 to 820 nm~\cite{27,28,29}, and have improved the precision of theoretical predictions.

Ions in Paul traps are well isolated from external fields and are thus widely used in precision metrology~\cite{30,31}, frequency standards~\cite{32,33,34,35,36}, quantum computing~\cite{37,38,39}, and quantum simulations~\cite{40,41,42}, etc. In our group, we are developing a $\mathrm{Yb^+}$ ions microwave clock~\cite{36}, which has advantages of miniaturization and integration. A 369 nm laser that drives the $\mathrm{^2S_{1/2}\rightarrow^2P_{1/2}}$ transition is needed in the $\mathrm{Yb^+}$ ions microwave clock to cool the ions and probe their states. Several methods could stabilize the 369 nm laser. One in particular locks it to a high-precision wavelength meter~\cite{43,44} that has a wide range for multi-wavelength locking. However, the wavelength meter can be very costly. The 369 nm laser could also be locked to  $\mathrm{Yb^+}$ ion spectroscopy via the plasma in a hollow cathode lamp~\cite{45}. However, the hollow cathode lamps usually have short lifetimes, which limits its application for long-term frequency locking. Another possibility is by locking the laser frequency to a low-drift Fabry-Perot cavity~\cite{46,47,48,49}, with a resulting long-term frequency drift less than 1 MHz~\cite{49} over several hours, if the cavity is installed in a vacuum chamber and is temperature stabilized. But the absolute frequencies of the locked lasers still cannot be determined directly, and the cost can be high. The 369 nm laser we used is a frequency-tripled laser. The seed laser is a 1108 nm distributed feedback (DFB) fiber laser, and it is power boosted by a fiber-optic power amplifier. The frequency of the 1108 nm laser is initially doubled via a periodically poled crystals (PPC) to generate 554 nm laser, which is then summed with the fundamental 1108 nm seed laser in another crystal set to generate 369 nm laser. Because the 554 nm and 369 nm outputs are generated from the same seed laser, the 554 nm laser could be stabilized via molecular iodine spectroscopy to obtain a frequency stabilized 369 nm laser for $\mathrm{Yb^+}$ ions applications. This method enables long-term frequency stability, low cost, and a compact volume.

In this work, we measured 13 transition lines of molecular iodine at 554 nm, and their hyperfine structures were obtained using modulation transfer spectroscopy (MTS). The absolute frequencies of selected hyperfine transition lines were determined with an optical frequency comb. We also investigated power and pressure shifts by varying the pump laser power and the temperature of the cold-finger in the iodine cell, respectively. Thus, the absolute transition frequency at zero-power and zero-pressure could be extrapolated. The R(146) 25-0 $a_{15}$ line of molecular iodine was used for $\mathrm{Yb^+}$ ions cooling, and the frequency of the stabilized 369 nm laser was approximately 44 MHz higher than the $\mathrm{^2S_{1/2}\rightarrow^2P_{1/2}}$ transition frequency of $\mathrm{Yb^+}$ ion. This frequency gap could be easy covered with an acoustical-optics modulator (AOM). The frequency stability of the 369 nm laser locked to the R(146) 25-0 $a_{15}$ line of $\mathrm{^{127}I_2}$ was $5 \times 10^{-12}$ over 1000 s. To the best of our knowledge, there were no precise measurements at 554 nm for molecular iodine, and our measurements could enrich high-precision experimental data, and also enable theoretical predictions. Meanwhile, the 369 nm frequency-tripled laser stabilized by molecular iodine spectroscopy also have wide applications in frequency metrology, and quantum information processing based on $\mathrm{Yb^+}$ ions.

\section{Experimental Setup}
The experimental setup is shown in {\bf Fig.~1}. It consists of the light source, the iodine frequency stabilization, and the optical frequency comb measurement.

\begin{center}
\includegraphics[width=0.99\linewidth]{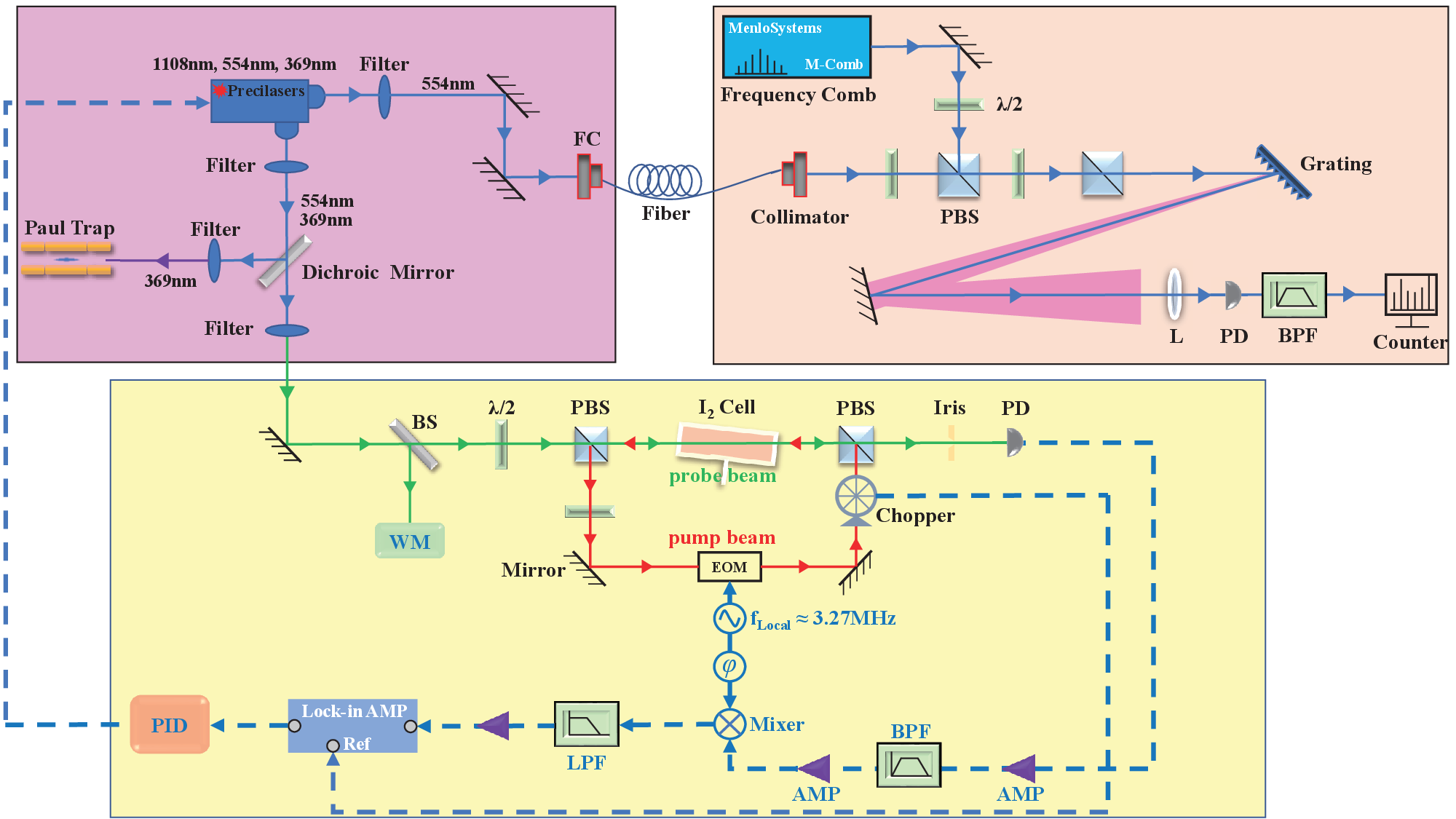}\\[5pt]  
\parbox[c]{15.0cm}{\footnotesize{\bf Fig.~1.} Experimental setup. FC, fiber coupler; PBS, polarization beam splitter; $\mathrm{\lambda/2}$, half-wave plate; L, lens; EOM, electro-optic modulator; Lock-in, lock-in amplifier; LPF, low-pass filter; BPF, band-pass filter; PD, photodetector; PID, electronic servo loop.}
\end{center}

The 369 nm frequency-tripled laser (PreciLasers, Inc.) has two outlets, and the output via both outlets contain the 1108 nm, 554 nm, and 369 nm lasers. The linewidth of the 369 nm laser is approximately 142 kHz, and the frequency could be tuned with an external modulation voltage. The 369 nm and 554 nm lasers from the main output port are separated via a dichroic mirror and optical filters. The rest 369 nm laser is directed to the ion trap, and its power is approximately 45 mW. The power of the rest 554 nm laser is approximately 24 mW, and it is used for measuring the molecular iodine spectroscopy via MTS. A small amount of 554 nm output is separated and filtered, and then coupled into an optical fiber for absolute frequency measurements with the optical frequency comb.

The MTS method for the 554 nm molecular iodine spectroscopy separates the 554 nm laser output with a polarization beam splitter (PBS) into two beams. One is the pump beam (approximately 15 mW), and the other is a probe beam (approximately 8 mW). The beam diameter is approximately 2 mm, and the two beams are overlapped and counter-propagate in the iodine cell. They are aligned to avoid laser feedback and scattering that could lead to strong noise.

The pump beam is phase-modulated with an electro-optic modulator (EOM) operated at frequency of 3.27 MHz. The probe beam is monitored with a fast photodetector (Thorlabs APD210), the output of which is mixed with the phase-adjustable local oscillator signal. The output of the mixer is filtered with a low-pass filter to remove high-order harmonics, and is amplified with a low-noise amplifier (LNA). To improve the signal-to-noise ratio (SNR) of the spectroscopy signal, a lock-in amplifier (Stanford Research Systems, Inc. SR830) is used in conjunction with a mechanical optical chopper (Stanford Research Systems, Inc. SR540) at 2 kHz. The iodine cell is fabricated by the Institute of Scientific Instruments, Academy of Sciences of the Czech Republic. It has a length of 10 cm and a diameter of 30 mm. The cold-finger temperature in the iodine cell is stabilized at $\mathrm{10.0 \pm 0.1^\circ C}$ to reduce pressure shifts and pressure broadening. A small part of the 554 nm laser is coupled to a precision wavelength meter (High Finesse WS-U) that has a frequency resolution of 1 MHz. The output of the lock-in amplifier is used to control the frequency of the seed laser via proportional integral derivative (PID) circuits.

To measure the absolute frequency of the 554 nm laser with the optical frequency comb (Menlo Systems GmbH, FC1500-250-ULN), the laser is locked to one of the iodine hyperfine transition components. The repetition rate of the optical frequency comb is approximately 250 MHz, and the offset frequency is stabilized at 35 MHz. Both the repetition rate and the offset frequency of the frequency comb are phase locked to a hydrogen maser (Microsemi, MHM 2010). The frequency of the hydrogen maser is measured against universal time coordinated (UTC) by the common view method via a global navigation satellite system (GNSS) receiver. The spectrum of the optical frequency comb covers a range of 500-1050 nm with a photonic crystal fiber (PCF) module. The measured laser and the optical frequency comb laser are combined into a photodetector with a PBS and a grating. The beat signal is filtered with a band-pass filter, and amplified with two cascaded LNAs. The beat frequency is measured by a frequency counter (KEYSIGHT 53230A) that is also referenced to the hydrogen maser. 

\section{Experimental Results}
\subsection{Observation of hyperfine components}
For the rigid diatomic iodine molecule, the selection rule for rotational transitions is $\triangle J = J^{\prime} - J^{\prime\prime} = \pm 1$, where $J$ is the rotational quantum number. The R branch corresponds to $\triangle J = +1$, and the P branch corresponds to $\triangle J = -1$. It has 15 hyperfine components when the ground state has an even rotational quantum number and 21 hyperfine components when the ground state has an odd rotational quantum number. In our experiment, we tune the laser frequency by ramping the external modulation voltage and observe the MTS signals. We characterize 13 hyperfine transitions of $\mathrm{^{127}I_2}$ near 554 nm, namely, the R(50) 22-0, P(46) 22-0, P(121) 24-0, P(69) 25-1, R(146) 25-0, R(147) 28-1, P(160) 26-0, P(102) 26-1, R(96) 23-0, R(49) 22-0, P(45) 22-0, P(92) 23-0, and R(72) 25-1 transitions, as shown in {\bf Fig.~2.} The detuning frequencies of the hyperfine transitions are measured with the wavelength meter. All these patterns are recorded when the cold finger is $\mathrm{10^\circ C}$, the pump power is 7.2 mW and the probe power is 8.0 mW.

\begin{figure}[H]
    \centering
    \subcaptionbox{\label{fig:figure951}}
    {\includegraphics[width=0.5\linewidth]{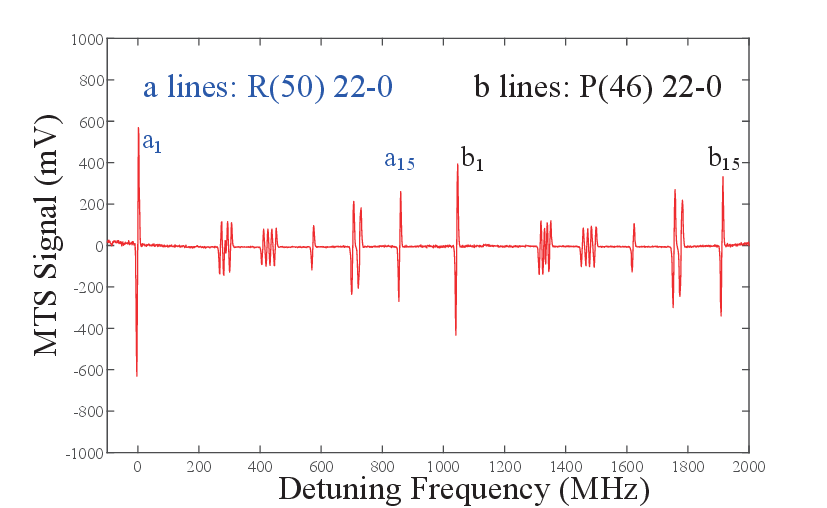}}\hfill
    \subcaptionbox{\label{fig:figure950}}
    {\includegraphics[width=0.5\linewidth]{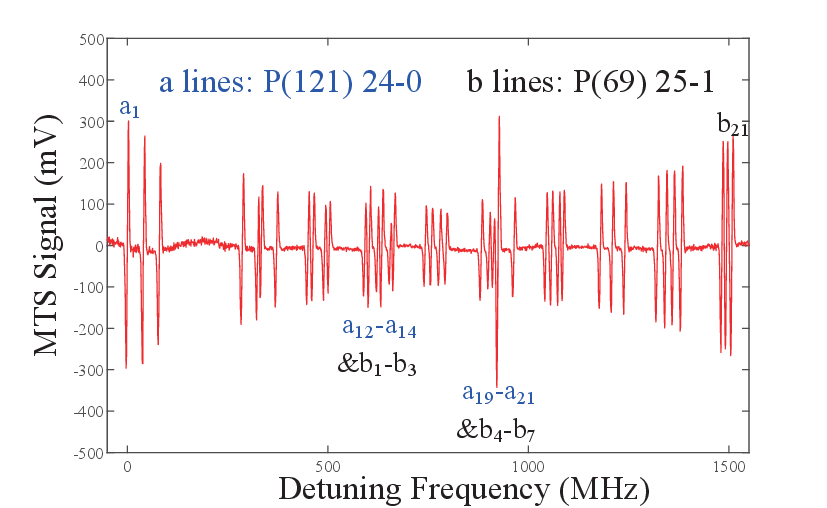}}\hfill
    \subcaptionbox{\label{fig:figure949}}
    {\includegraphics[width=0.5\linewidth]{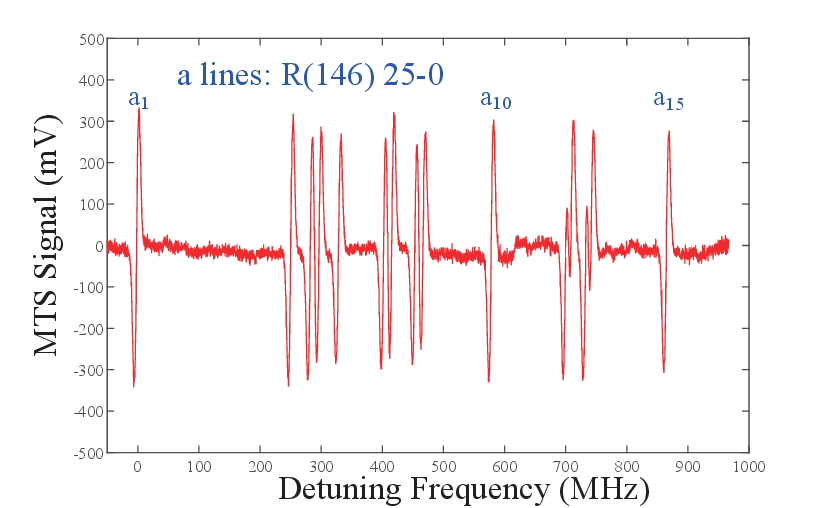}}\hfill
    \subcaptionbox{\label{fig:figure948-2}}
    {\includegraphics[width=0.5\linewidth]{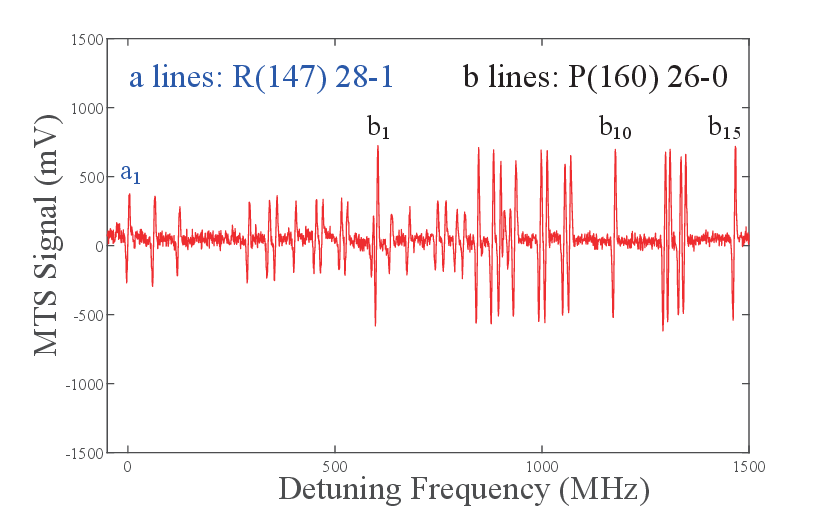}}\hfill
    \subcaptionbox{\label{fig:figure948-1}}
    {\includegraphics[width=0.5\linewidth]{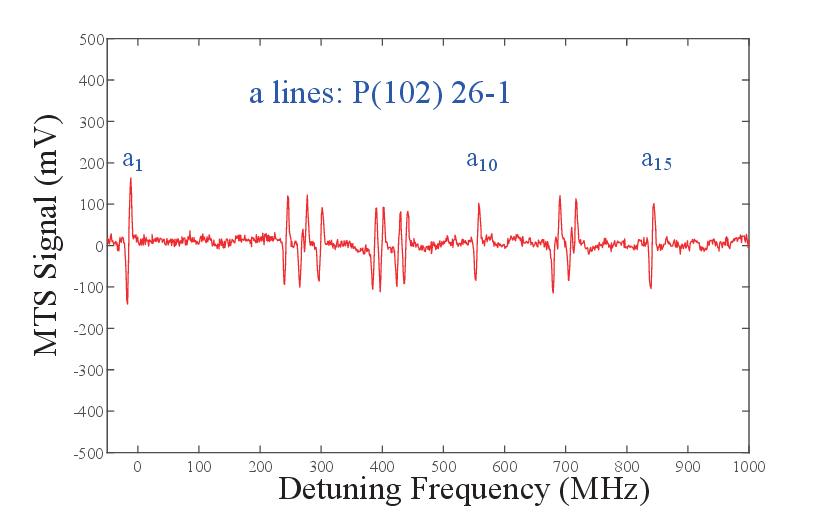}}\hfill
    \subcaptionbox{\label{fig:figure947}}
    {\includegraphics[width=0.5\linewidth]{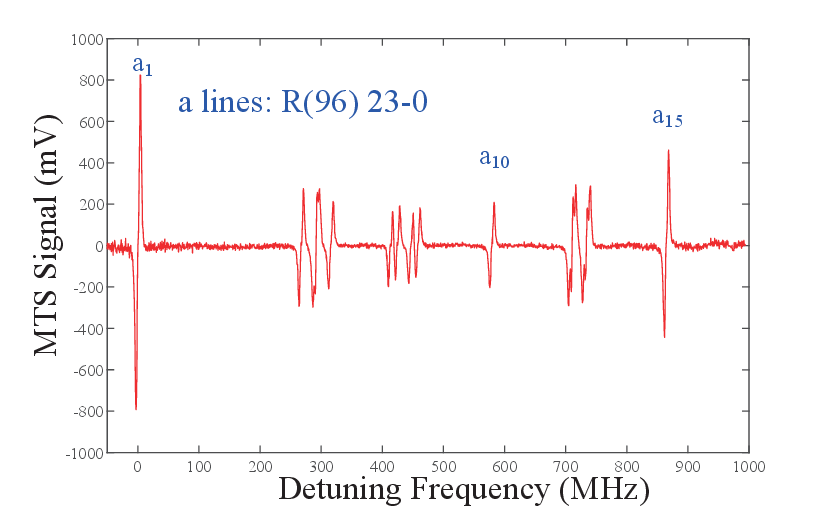}}\hfill
    \subcaptionbox{\label{fig:figure946}}
    {\includegraphics[width=0.5\linewidth]{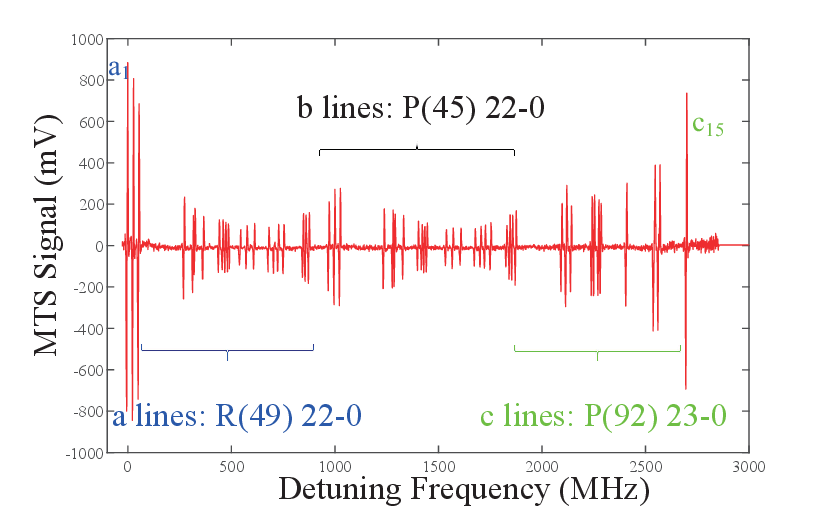}}\hfill
    \subcaptionbox{\label{fig:figure945}}
    {\includegraphics[width=0.5\linewidth]{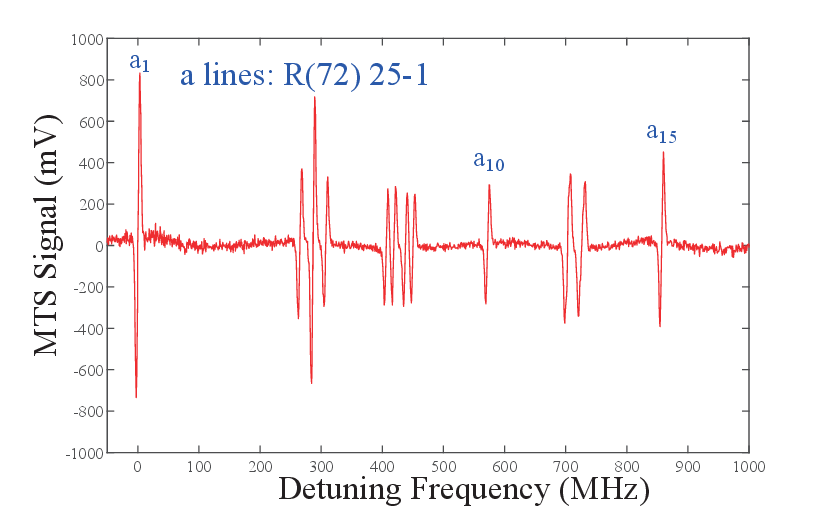}}\hfill
\parbox[c]{15.0cm}{\footnotesize{\bf Fig.~2.} Modulation transfer spectroscopy signals of $\mathrm{^{127}I_2}$ hyperfine transitions around 554 nm.}
\end{figure}

\subsection{Absolute frequency measurements}

The optical frequency comb is locked to the hydrogen maser to measure absolute frequencies. The laser frequency is measured by recording its beat frequency $f_b$ with the nearest frequency comb line: $$ f_{laser} = N\cdot f_r\pm f_o\pm f_b , $$ where $f_r$ is the repetition rate at approximately 250 MHz, $f_o$ is the 35 MHz offset frequency, and the mode number $N$ is determined by changing the repetition rate of the optical frequency comb in a two-step process without using a wavelength meter~\cite{50}.
 
As noted above, we determine the effects of pressure and power shifts to accurately determine the transition frequency at zero-pressure and zero-power. 

The iodine vapor pressure is a function of the cold finger temperature in
the cell~\cite{51}: 
$$ \mathrm{log}(P) = \frac{-3512.830}{T} - 2.013\cdot \mathrm{log}(T) + 18.37971, $$
where $P$ is the iodine vapor pressure in pascals, and $T$ is the cold finger temperature in kelvin. The pump power is fixed at 7.2 mW, and the probe power is fixed at 8.0 mW. We vary the cold finger temperature to change the vapor pressure from 2.6 Pa to 3.9 Pa. Typical data of the laser locked to the P(69) 25-1 $b_{18}$ line are shown in {\bf Fig.~3.}, which indicates good linearity between the iodine vapor pressure and the hyperfine transition frequency of the P(69) 25-1 $b_{18}$ line of $\mathrm{^{127}I_2}$. By linear fitting, we obtain a pressure shift coefficient of $\mathrm{-78.2 \pm 8.8}$ kHz/Pa which is used to correct the measured results to zero-pressure values of the P(69) 25-1 $b_{18}$ line.

\begin{figure}[H]
    \centering
    \includegraphics[width=0.9\textwidth]{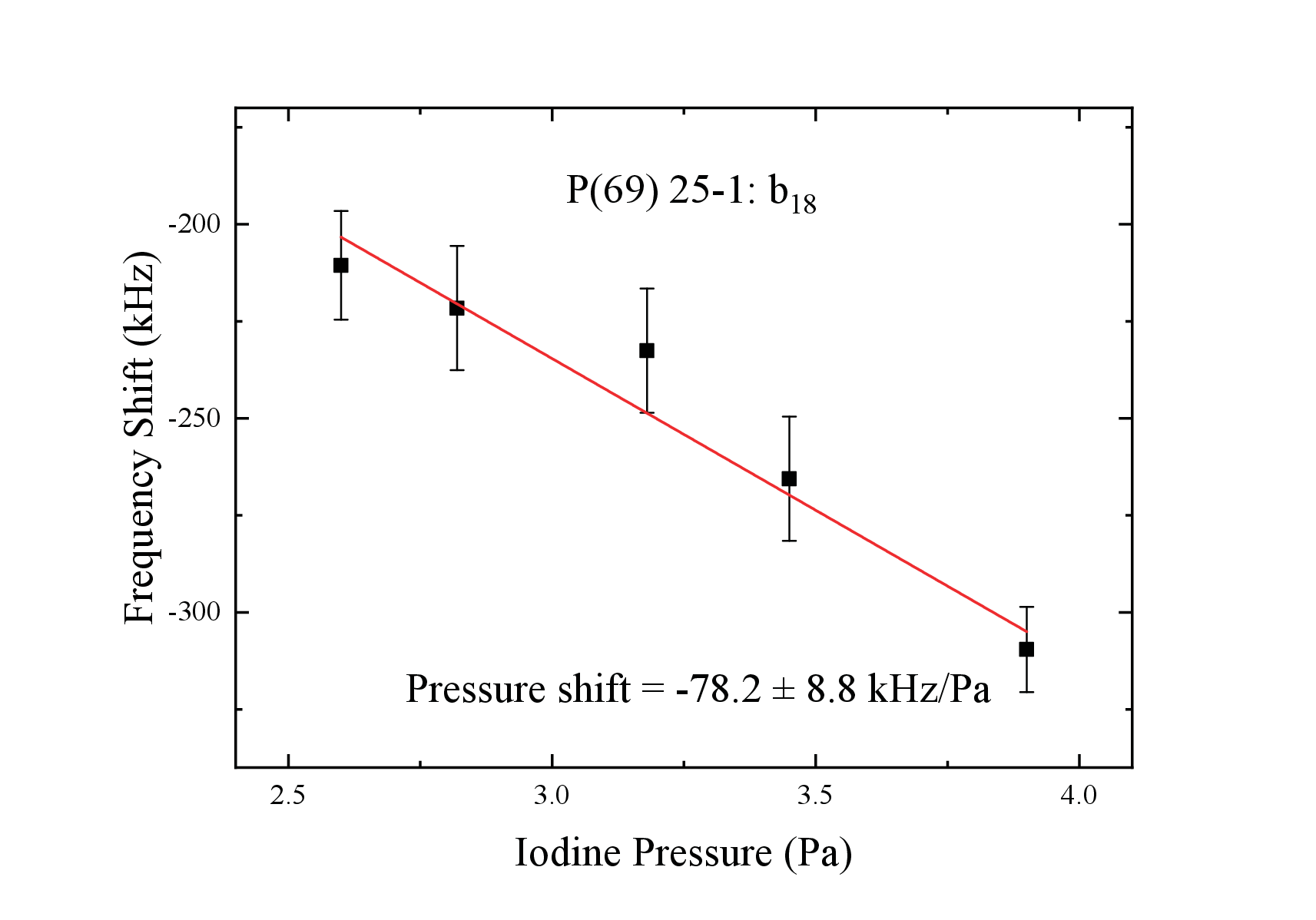}
    \parbox[c]{15.0cm}{\footnotesize{\bf Fig.~3.} (Color online) Pressure shift of the 554 nm laser locked to the P(69) 25-1 $b_{18}$ line of $\mathrm{^{127}I_2}$. The pump power is fixed at 7.2 mW, and the probe power is fixed at 8.0 mW. Each point is an average of ten measurements. The solid red line is a linear fit.}
\end{figure}

To investigate the power shift, the cold finger temperature is fixed at $\mathrm{10^\circ C}$, corresponding to an iodine vapor pressure of 2.8 Pa, the probe power fixed at 8.0 mW, and the pump power is varied from 5.0 mW to 7.2 mW. Typical data of the laser locked to the P(69) 25-1 $b_{18}$ line are shown in {\bf Fig.~4.}, which indicates good linearity between the pump power and the hyperfine transition frequency of the P(69) 25-1 $b_{18}$ line of $\mathrm{^{127}I_2}$. By linear fitting, we obtain a power shift coefficient of $\mathrm{17.9 \pm 0.9}$ kHz/mW, which is used to correct the measured results to zero-power values of the P(69) 25-1 $b_{18}$ line.

\begin{figure}[H]
    \centering
    \includegraphics[width=0.9\textwidth]{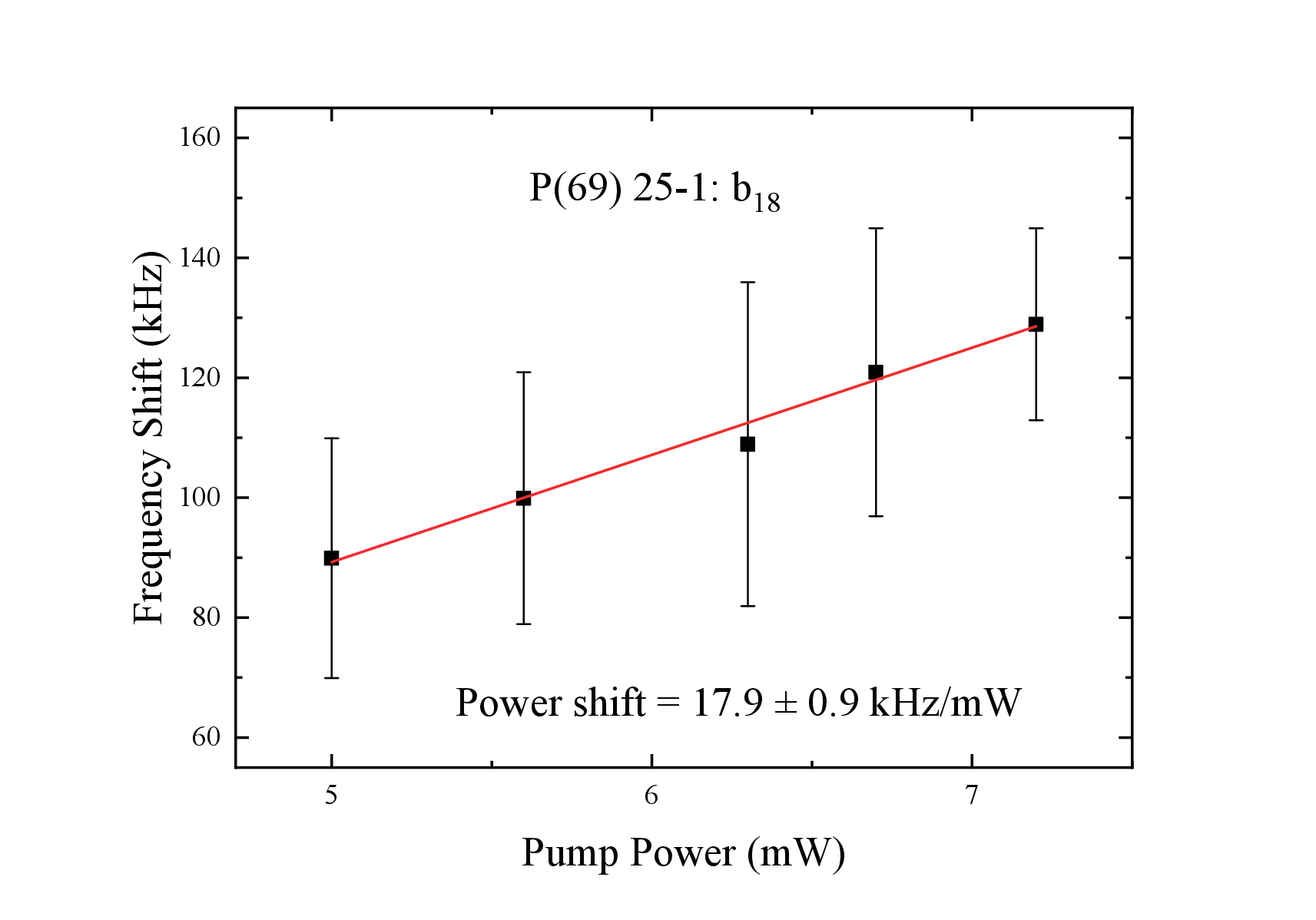}
    \parbox[c]{15.0cm}{\footnotesize{\bf Fig.~4.} (Color online) Power shift of the 554 nm laser locked to the P69 (25-1) $b_{18}$ line of $\mathrm{^{127}I_2}$. The cold finger temperature is fixed at $\mathrm{10^\circ C}$ (2.8 Pa). The probe power is fixed at 8.0 mW. Each point is an average of ten measurements. The solid red line is a linear fit.}
\end{figure}

According to the above measurements and analyses, the total systematic uncertainty (Type-B) is 30 kHz. All uncertainties are listed in {\bf Table 1.} Thus, we are able to use these coefficients to obtain the zero-pressure and zero-power values of the P(69) 25-1 $b_{18}$ line of $\mathrm{^{127}I_2}$ . After the calculation, the measured frequency is 90 kHz smaller than the absolute frequency at zero-pressure and zero-power. Therefore, we add this deviation to the measured frequencies.

\begin{center}
{\footnotesize{\bf Table 1.} Uncertainties in the measured frequency of the P(69) 25-1 $b_{18}$ line.\\
\vspace{2mm}
\begin{tabular}{ccc}
\hline
{Effect}      &{shift (kHz)}        & {Uncertainty (kHz)} \\
\hline
{Pressure shift}    & {219.0}    &{29.4}\\
{Power shift}     & {-128.9}    &{5.8}\\
{Total type-B uncertainty}  & {-}    &{30}\\
{Type-A uncertainty}  & {-}    &{33}\\
{Total uncertainty}  & {-}    &{44}\\
\hline
\newline
\end{tabular}}
\end{center}

Although the pressure and power shifts are not exactly the same for different hyperfine transition lines of $\mathrm{^{127}I_2}$, the coefficients of the close transitions are normally similar within the same order of magnitude~\cite{12,15}. We use the pressure and power shift coefficients of the P(69) 25-1 $b_{18}$ line as a reference to evaluate frequency shifts of other measured hyperfine transition lines of $\mathrm{^{127}I_2}$ and to make rough corrections. The final corrected results of the measured transition frequencies are listed in {\bf Table 2.} These measurements cover a frequency range of approximately 100 GHz around 554 nm.

\begin{center}
{\footnotesize{\bf Table 2.} Results of selected hyperfine transition frequencies.\\
\vspace{2mm}

\begin{threeparttable}
\setlength{\tabcolsep}{15mm}{
\begin{tabular}{cc}
\hline
{Hyperfine component}      & {Measured$^a$(kHz)} \\
\hline
{R(50) 22-0} $a_1$ line   & {540 844 838 191 (46)(30)}\\
{P(121) 24-0} $b_{1}$ line   & {540 847 867 829 (54)(30)}\\
{P(69) 25-1} $b_{18}$ line   & {540 849 384 667 (33)(30)}\\
{R(146) 25-0} $a_{15}$ line   & {540 859 322 747 (48)(30)}\\
{P(160) 26-0} $a_{15}$ line   &  {540 861 050 549 (58)(30)}\\
{P(102) 26-1} $a_1$ line   & {540 864 607 341 (26)(30)} \\
{R(96) 23-0} $a_{10}$ line   & {540 868 830 629 (47)(30)} \\
{P(92) 23-0} $a_{15}$ line   &  {540 883 006 518 (72)(30)}\\
{R(72) 25-1} $a_{10}$ line   &  {540 886 126 160 (41)(30)}\\
\hline
\end{tabular}}
    \begin{tablenotes}
        \footnotesize
        \item $^a$Type-A uncertainties (k=2) are given in the first parentheses, and type-B uncertainties are given in the second parentheses.
      \end{tablenotes}
\end{threeparttable}}
\end{center}

\subsection{Frequency Stability}
We lock the laser to the R(146) 25-0 $a_{15}$ line of $\mathrm{^{127}I_2}$  because its corresponding frequency is very close to that of $\mathrm{^2S_{1/2}\rightarrow^2P_{1/2}}$ transition of the $\mathrm{^{171}Yb^+}$ ions (offset by approximately 44 MHz). The demodulated error signal of the lock-in amplifier is fedback to the seed laser with the electronic servo system. The frequency stability of the laser locked to the R(146) 25-0 $a_{15}$ line of $\mathrm{^{127}I_2}$ is measured by recording the beat frequency between the iodine-stabilized laser and the optical frequency comb over a 1 s gate time. The total measurement time is approximately 16000 s. The calculated Allan deviation is shown in {\bf Fig.~5.}, together with the Allan deviation of the free-running laser. The stability of the iodine-stabilized laser reaches a level of $5 \times 10^{-12}$ over a 1000 s integration time. After improving its stability and reproducibility, this iodine-stabilized laser could be used as a reference laser for the $\mathrm{^2S_{1/2}\rightarrow ^2P_{1/2}}$ transition of the $\mathrm{Yb^+}$ ions.

\begin{figure}[H]
    \centering
    \includegraphics[width=0.95\textwidth]{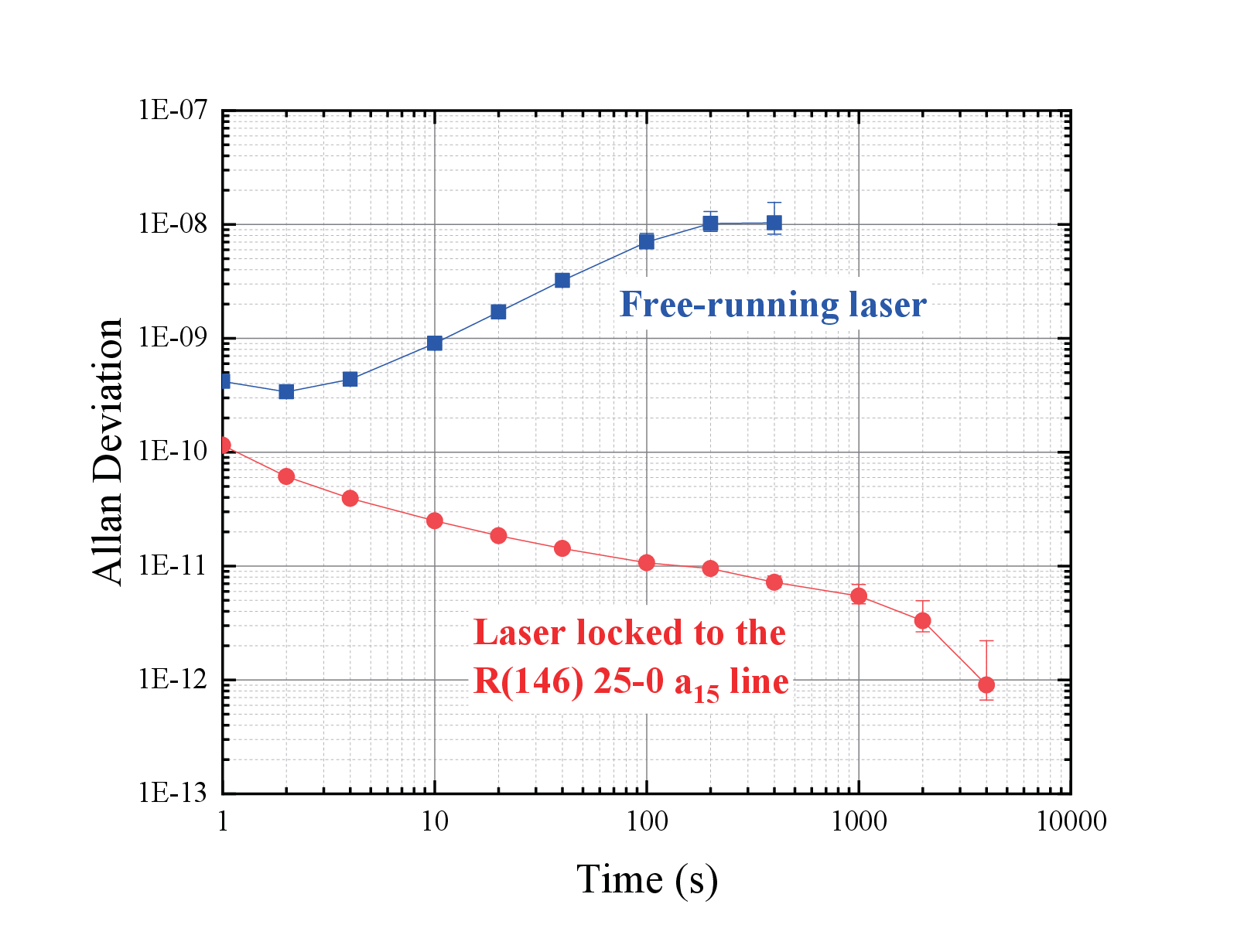}
    \parbox[c]{15.0cm}{\footnotesize{\bf Fig.~5.} (Color online) Curve with red circles: Allan deviation of the measured beat frequency between the laser that is locked to the R(146) 25-0 $a_{15}$ line and the frequency comb that is locked to the hydrogen maser. Curve with blue squares: Allan deviation of the free-running laser.}
\end{figure}

\section{Conclusions}
In conclusion, we have investigated 13 hyperfine transitions of $\mathrm{^{127}I_2}$ near 554 nm, namely, the R(50) 22-0, P(46) 22-0, P(121) 24-0, P(69) 25-1, R(146) 25-0, R(147) 28-1, P(160) 26-0, P(102) 26-1, R(96) 23-0, R(49) 22-0, P(45) 22-0, P(92) 23-0, and R(72) 25-1 transitions, and the transition frequencies of the hyperfine components have been measured via MTS along with an optical frequency comb. The measurements provide good guidance for further improvements in theoretical calculations at approximately 554 nm. The frequency stability of the iodine-stabilized laser locked to the R(146) 25-0 $a_{15}$ line is $5 \times 10^{-12}$ over a 1000 s integration time. The iodine-stabilized 369 nm laser could be used for atomic clocks and quantum information processing based on $\mathrm{Yb^+}$ ions.

\addcontentsline{toc}{chapter}{Acknowledgment}
\section*{Acknowledgments}
This work is supported by National Natural Science Foundation of China (12073015) and National Key Research and Development Program of China (2021YFA1402100).

\end{document}